\documentclass[prl,twocolumn,showpacs,floatfix]{revtex4}
\usepackage{epsfig,amssymb,amscd,amsmath, graphicx}
\usepackage{subfigure}

\newcommand{\nin}{\noindent}

\newcommand{\bea}{\begin{eqnarray}}  
\newcommand{\eea}{\end{eqnarray}} 
\newcommand{\be}{\begin{equation}}  
\newcommand{\ee}{\end{equation}}  

\newcommand{\ket}[1]{ | \, #1  \rangle} 
\newcommand{\bra}[1]{ \langle #1 \,  |}
\newcommand{\ZZ}{\mathbb{Z}}

\begin{document}

\title{Topological degeneracy and vortex manipulation in Kitaev's honeycomb model}

\author{G. Kells$^{1}$, A. T. Bolukbasi$^{1}$, V. Lahtinen$^{2}$,
J. K. Slingerland$^{3}$, J. K. Pachos$^{2}$ and J. Vala$^{1,3}$}


\affiliation{$^{1}$ Department of Mathematical Physics, National
University of Ireland, Maynooth, Ireland, \\ $^{2}$ School of Physics
and Astronomy, University of Leeds, Leeds LS2 9JT, UK, \\ $^{3}$
Dublin Institute for Advanced Studies, School of Theoretical Physics,
10 Burlington Rd, Dublin, Ireland. }

\begin{abstract}
The classification of loop symmetries in Kitaev's honeycomb lattice
model provides a natural framework to study the abelian topological
degeneracy. We derive a perturbative low-energy effective Hamiltonian,
that is valid to all orders of the expansion and for all possible
toroidal configurations. Using this form we demonstrate at what order
the system's topological degeneracy is lifted by finite size effects
and note that in the thermodynamic limit it is robust to all
orders. Further, we demonstrate that the loop symmetries themselves
correspond to the creation, propagation and annihilation of
fermions. Importantly, we note that these fermions, made from pairs of
vortices, can be moved with no additional energy cost.
\end{abstract}

\pacs{05.30.Pr, 75.10.Jm, 03.65.Vf}

\date{\today}
\maketitle

Recently, Kitaev introduced a spin-1/2 quantum lattice model with
abelian and non-abelian topological phases \cite{Kit06}. This
model is relevant to on-going research into
topologically fault-tolerant quantum information processing
\cite{Kit03,Pac05,Nay07}. The system comprises of two-body
interactions and is exactly solvable, which makes it attractive both
theoretically
\cite{Pac06,Zha06,Chen07a,Chen07b,Bas07,Sch07,Lah07,Feng07,Vid08,Zha08,Dus08,Woo08}
and experimentally \cite{Dua03,Mic06,Jia07,Agu08}.

Here, by classifying the loop symmetries of the system according to
their homology, we derive a convenient form of the effective
Hamiltonian on the torus. The result is valid for all orders of the
Brillouin-Wigner perturbative expansion around the fully dimerized
point as well as for all toroidal configurations. This allows the system's
topological degeneracy to be addressed and shows at what order in the
expansion the degeneracy is lifted. In the thermodynamic limit the
system's topological degeneracy remains to all orders. In a separate
analysis, valid for the full parameter space, we examine the
paired-vortex excitations created by applying certain open string
operations to the ground state. These vortex-pairs are fermions and
can be freely transported in a way that keeps additional unwanted excitations to a minimum.

\begin{figure}
       \includegraphics[width=.27\textwidth,height=0.15\textwidth]{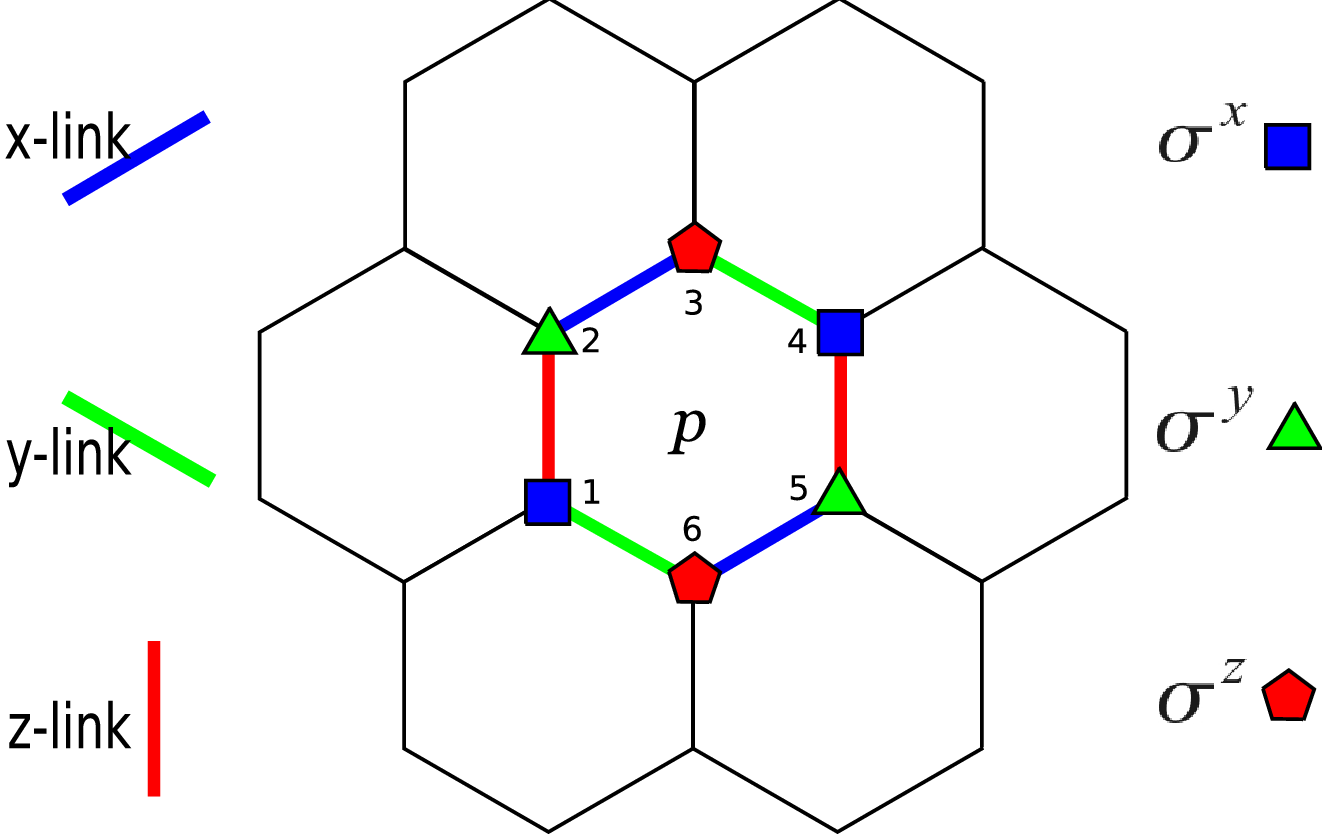}
       \caption{ The honeycomb lattice and the operator $W_p$.} 
       \label{fig:fulllattice}
\end{figure}

The Hamiltonian for the system can be written as
\be
H  = - \sum_{\alpha \in \{ x,y,z \}} \sum_{i,j} J_\alpha K_{ij}^{\alpha,\alpha}
\label{eq:H}
\ee \nin where $K_{ij}^{\alpha,\beta} \equiv \sigma_i^\alpha \otimes
\sigma_j^\alpha$ denotes the exchange interaction occurring between
the sites ${i,j}$ connected by a $\beta$-link, see
FIG. \ref{fig:fulllattice}. In what follows we will use
$K^{\alpha}_{ij} \equiv K^{\alpha,\alpha}_{ij}$ whenever
$\alpha=\beta$. Following \cite{Kit06}, we consider loops of $n$
non-repeating $K$ operators, $K^{\alpha^{(1)}}_{ij}
K^{\alpha^{(2)}}_{jk} ,......, K^{\alpha^{(n)}}_{li}$, where
$\alpha^{(m)} \in {x,y,z}$. Any loop constructed in this way commutes
with the Hamiltonian and with all other loops. When the model is
mapped to free Majorana fermions coupled to a $\ZZ_2$ gauge field,
these loop operators become Wilson loops \cite{Kit06}.
The plaquette operators
\be
W_p = \sigma_1^x \sigma_2^y \sigma_3^z  \sigma_4^x \sigma_5^y \sigma_6^z,
\label{eq:Wp}
\ee
where the numbers $1$ through $6$ label lattice sites on single
hexagonal plaquette $p$, see FIG. \ref{fig:fulllattice}, are the
closed loop operators around each of the hexagons of the
lattice. The commutation relations imply that we may choose energy eigenvectors $\ket{n}$ such that $w_{p}=\bra{n}
W_p \ket{n}=\pm 1$.  If $w_{p} = -1$, we say that the state
$\ket{n}$ carries a vortex at $p$.

For a finite system of $N$ spins on a torus there are $N/2$
plaquettes. The product of all plaquette operators is the identity and
this is the single nontrivial relation between them. Hence there are
only $N/2 -1$ independent quantum numbers, $\{ w_1, ...., w_{N/2 -1}
\}$. All homologically trivial loops are products of plaquettes. The
relevant homology is $\ZZ_2$, since loop operators square to
the identity. To describe the full symmetry group generated by loop
operators, we introduce generators for the nontrivial $\ZZ_2$ homology
classes of the surface that the lattice lives on. At most one
generator per homology class is necessary, since all elements of any
homology class can be generated from an arbitrary element of that
class using the plaquettes. The $\ZZ_2$ homology group of the torus is
$\ZZ_2\times\ZZ_2$, so it is enough to add two homologically
nontrivial loops as generators. The third nontrivial class is
generated from the product of these two. The full loop symmetry group
of the torus is the abelian group with $N/2+1$ independent generators
of order $2$, that is $\ZZ_{2}^{N/2+1}$. All closed loop symmetries
can be written as
\be
\label{eq:genloop}
C_{(k,l)}=G_k F_l(W_1,W_2,.....,W_{N-1}).  
\ee 
Here $k \in\{0,1,2,3\}$, $G_0 = I$ and $G_1$, $G_2$ and $G_3$ are arbitrarily
chosen symmetries from the three nontrivial homology classes. The
$F_{l}$, with $l\in\{1,\ldots,2^{N/2-1}\}$, run through all monomials
in the $W_{p}$.

The loop symmetries play an important role in the perturbation theory
of the abelian phase of the model. Following Kitaev we take $J_z \gg
J_x,J_y$ and write the Hamiltonian as $H=H_0+U$, where $H_0= -J_z
\sum_{ij} K_{ij}^z$ is the unperturbed Hamiltonian and $U =-
\sum_{\alpha \in \{ x,y \}} J_\alpha \sum_{ij} K_{ij}^{\alpha}$ is the
perturbative contribution. $H_0$ has a $2^{N/2}$-fold degenerate
ground state space spanned by ferromagnetic configurations of the dimers on
$z$-links. To understand how this degeneracy behaves under
perturbation we analyse the Brillouin-Wigner expansion
\cite{Zim69,Kill77}. The method returns the systems energies $E$ as an
implicit non-linear eigenvalue problem and thus, for the actual
calculation of coefficients to high orders, can be difficult to apply
\cite{Berg07}.  However, we will take advantage of the infinite but
exact nature of the series by recognizing that on the torus the form
of the Hamiltonian is restricted, allowing one term for each element
of the group of loop symmetries. This will facilitate a general
discussion on the system's topological degeneracy.

Define $\mathcal{P}$ to be the projector onto the ferromagnetic subspace and note that
for any exact eigenstate of the full Hamiltonian $\ket{\psi}$, the
projection $\ket{\psi_0}=\mathcal{P}\ket{\psi}$  satisfies 
\be \left[E_0 +
\sum_{n=1}^\infty H^{(n)} \right] \ket{\psi_0} =E \ket{\psi_0}=
H_{\text{eff}} \ket{\psi_0} ,
\label{eq:SEF}
\ee 
where $H^{(n)} = \mathcal{P} U \mathcal{G}^{n-1}
\mathcal{P}$, $\mathcal{G}=\left[1/(E-H_0)\right](1-\mathcal{P}) U $
and $E_0$ is the ground state energy of $H_0$. The eigenstates, with eigenvalue $E$, of the effective system and full system are related by the expression
$\ket{\psi}=(1-\mathcal{G})^{-1}\ket{\psi_0}=\sum_{n=0}^\infty
\mathcal{G}^n \ket{\psi_{0}}$.

Calculating the $n^{\mathrm{th}}$ order correction is equivalent to
finding the non-zero elements of the matrix $H^{(n)}$. Contributions
to $H^{(n)}$ come from the length $n$ products $K^{\alpha^{(1)}}_{ij} 
\ldots, K^{\alpha^{(n)}}_{kl}$ with $\alpha^{(m)} \in {x,y}$ that
preserve the low-energy subspace. Hence any such contribution comes
from an element of the group of loop symmetries from which all factors
$K_{ij}^{z}$ have been removed.

\begin{figure}
       \includegraphics[width=.38\textwidth,height=0.2\textwidth]{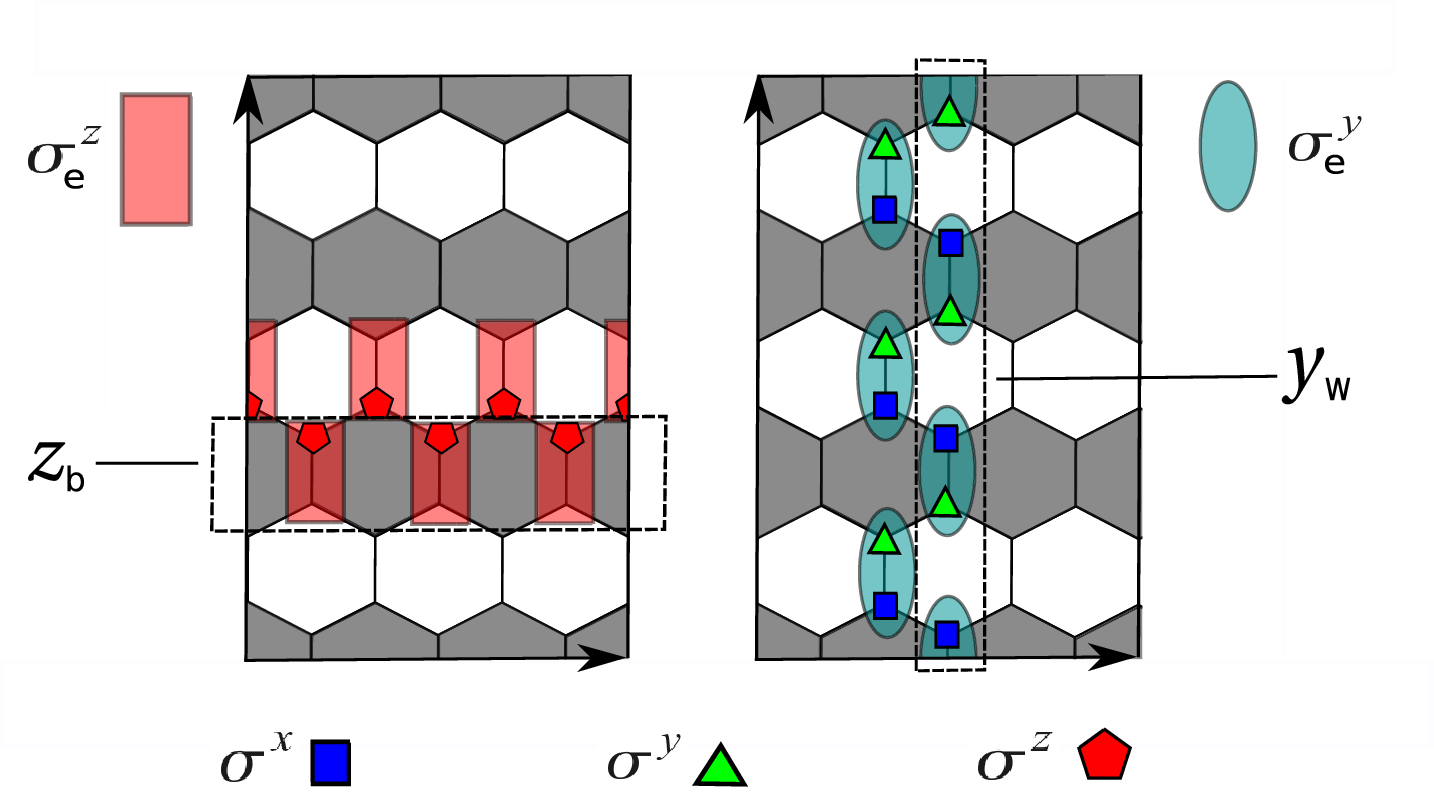}
       \caption{ The $Z$ and $V$ chains with their projections on to
       the dimerized subspace. The projections may be factorised into
       products $\mathcal{P}[Z]\rightarrow z_b z_w$ and
       $\mathcal{P}[V] \rightarrow y_b y_w$. Each of the individual
       factors $z_b,z_w,y_b,y_w$ also commute with the homologically
       trivial components of the effective Hamiltonian but obey the
       relation $z_{j}^{-1} y_{k}^{-1} z_j y_k = e^{i \pi
       (1-\delta_{jk})}I$. In the text, dimers are referred to as
       black (white) if they are shared by a black (white) plaquette.}
       \label{fig:effective}
\end{figure}

The resulting low-energy effective Hamiltonian can be written in terms
of operators acting on the spins of the dimers using the following
transformation rules:
\begin{equation}
\begin{array}{rclrcl}
\mathcal{P}[\sigma^{x}\otimes\sigma^{y}] & \rightarrow & +\sigma_{\text{e}}^{y}, &
\mathcal{P}[\sigma^{x}\otimes\sigma^{x}] & \rightarrow & +\sigma_{\text{e}}^{x}, \\
\mathcal{P}[\sigma^{y}\otimes\sigma^{y}] & \rightarrow & -\sigma_{\text{e}}^{x}, &
\mathcal{P}[\sigma^{z}\otimes I \;\; ] & \rightarrow & +\sigma_{\text{e}}^{z}, \\
\mathcal{P}[\sigma^{z}\otimes\sigma^{z}] & \rightarrow & +I_{\text{e}}, &
  \\
\end{array}
\end{equation}
where the subscript $\text{e}$ indicates the effective spin
operation and the arrow $\rightarrow$ can be read as 'is represented by'. Importantly, this transformation can be applied directly to the loop
symmetries themselves, without removing the $z$-links first, and does
not change the resulting operator on the low-energy subspace. The lowest order non-constant contributions therefore
come from the plaquette operators $\mathcal{P} [W_p] \rightarrow Q_p =
\sigma_{e(l)}^y \sigma_{e(r)}^y \sigma_{e(u)}^z \sigma_{e(d)}^z$,
where $l,r,u,d$ denotes the positions (left, right, up and down) of
the effective spins, relative to the plaquette $p$
\cite{Kit06}. Expanding to all orders, we have contributions from all
loop symmetries, both homologically trivial and non-trivial. To come
to an explicit expression for the effective Hamiltonian, we now
introduce a particular generating set for the loop symmetry group,
constructed from $N/2-1$ plaquettes and the operators $Z \equiv
\prod_i \sigma_i^{z}$,where $i$ represents lattice sites in the
horizontal direction of alternating $x$ and $y$-links and $V \equiv
\prod K_{jk}^{x,y} \prod K_{lm}^{y,x}$, where the products take place
over vertically arranged $x$- and $y$-links. The projections
$\mathcal{P}(Z) \rightarrow z $ and $\mathcal{P}(V) \rightarrow y$ act
by $\sigma_{e}^{z}$ and $\sigma_{e}^{y}$ on the relevant effective
spins, see FIG. \ref{fig:effective}. In analogy to (\ref{eq:genloop})
we can now write the full effective Hamiltonian as
\be \label{eq:Heff}
H_{\text{eff}} = \sum_{k=0}^3 \sum_{l=1}^{2^{N/2-2}} d_{k,l} G_k(z,y)  F_l(Q_{1},Q_{2},...,Q_{N/2-2}),
\ee
\nin where $G_0 = I, G_1=z, G_2=y$, $G_3=z y$ and the $d_{k,l}$ are
constants which depend on $J_{x}$, $J_{y}$ and $J_{z}$. This form is
strictly valid for when the effective square toroidal lattice has an
even number of plaquettes $Q_p$ along both directions. The inside sum
only runs to $2^{N/2-2}$ because, as a result of the projection, we
now have two non-trivial relations $\prod Q_b =\prod Q_w =1$, see
FIG. \ref{fig:effective}. The arguments here apply to even-even toric code configurations but can be generalised to cover the odd-odd and odd-even lattices examined in \cite{Wen03}.

In general, $d_{k,l} \sim O \left( J^{n_{x}}_{x} J^{n_{y}}_{y} \right)
$, where $n_x$ and $n_y$ are the respective number of x-links and
y-links used to make $ G_k(z,y) F_l(Q_{1},Q_{2},...,Q_{N/2-2})$.  In
the thermodynamic limit, and for homologically nontrivial loops
$(k>0)$, the values of $n_x$ and $n_y$ go to infinity and the limiting
form of (\ref{eq:Heff}) is similar to the planar Hamiltonian addressed
in \cite{Sch07} but with additional topological degrees of freedom.

We can now analyse the topological degeneracy of the abelian
phase. The general argument for topological ground state degeneracy
depends on the existence of operators $T_1$ and $T_2$ that both create
particle/anti-particle pairs from the vacuum, bring the particle (or
anti-particle) around the torus and then annihilate the pair
\cite{Ein90,Kit03}. These operators should commute with the
Hamiltonian but not with each other. Hence $T_1$ and $T_2$ operators
for the honeycomb system cannot be contained within the group of
commuting loop symmetries. However, the low-energy effective
representations of the homologically nontrivial loops' generators have
the factorizations $ z=z_b z_w$ and $y=y_b y_w$, where $z_b$ and $y_b$
act with effective $\sigma^z$'s and $\sigma^{y}$'s respectively on the
spins of the `black' dimers involved in $z$ and $y$, while $z_w$ and
$y_w$ do the same for the `white' dimers, (see
FIG. \ref{fig:effective}).

These 'black' and 'white' operators correspond to the nontrivial loop
operators on the square lattice and dual square lattice of the toric
code (cf.~\cite{Kit03}) and thus obey the commutation relations
$z_{j}^{-1} y_{k}^{-1} z_j y_k = e^{i \pi (1-\delta_{jk})}I$. Since
these operators commute with the effective plaquette operators $Q_p$
they also commute with all homologically trivial components of
$H_{\text{eff}}$.  However, they do not commute with all of the
homologically non-trivial components. If we define $C'$ as the
homologically non-trivial loop with the least number of x- and
y-links, then the topological degeneracy is first broken at the order
$M$, where $M$ is the number of x- and y-links in $C'$.

At any size, the plaquette and homologically non-trivial operators together generate all conserved quantities and in particular, determine the energy. For the typical system sizes that can be handled by numerical diagonalization and other numerical methods, the homologically non-trivial terms in the effective Hamiltonian are appreciable and must be taken into account to produce a good fit to exact numerical results. In larger tori these homologically non-trivial terms become less relevant to the energy and the topological degeneracy of
the system can be robust beyond the 4th order toric code
approximation. Indeed, in the thermodynamic limit, the 4-fold topological degeneracy exists to all
orders of the perturbation theory and the
eigenstates of the effective Hamiltonian are exactly those of the toric code. One should note however that even in this limit, and unlike the toric code, the energy of a particular eigenstate is also determined by the relative positions of the vortex excitations \cite{Sch07}. 

We now concentrate on the full Hamiltonian and consider the physical
properties associated with open ended strings of overlapping
$K^{\alpha}$ operators. We first note that $\{\sigma_i^{\alpha}, W_p
\}=0$ when the site $i$ belongs to an $\alpha$-link at plaquette
$p$. Hence, the operator $\sigma_j^{\alpha}$ changes the vorticity of
the two plaquettes sharing this $\alpha$-link by either creating or
annihilating a pair of vortices, or moving a vortex from one plaquette
to the other. It follows that the $K$ operators satisfy $[
K^{\alpha}_{ij}, W_p ] = 0 ~(\forall i,j)$, $[K^{\alpha,\beta}_{ij},
W_p ] = 0 ~(i, j \notin p$) and $ \{ K^{\alpha,\beta}_{ij}, W_p \} = 0
~( i \vee j \in p)$.

\begin{figure}
       \includegraphics[width=.3\textwidth,height=0.2\textwidth]{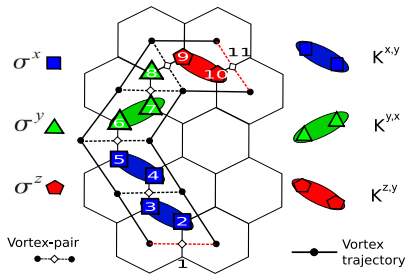}      
       \caption{ The operator $K^{x,y}_{2,3}$ is used to create two
       vortex-pairs from the vacuum with an energy cost of $2J_z
       \langle K_{1,2}^{z} + K_{3,4}^{z} \rangle$. The subsequent
       operators $K^{x,y}_{4,5}$ and $K^{y,x}_{6,7}$ move one of the
       pairs in the direction shown. The Pauli operator $\sigma^y_8$
       rotates this pair and the energy of the system at this time is
       $E_0 + 2 J_z\langle K_{1,2}^{z}\rangle + 2 J_x \langle
       K_{8,9}^{x} \rangle$. This new pair is then moved horizontally
       with no additional energy cost by $K_{9,10}^{z,y}$.}
       \label{fig:movement}
\end{figure}

Now define a path $s$ on the lattice as some ordered set of $|s|$
neighboring sites connecting the endpoints $i$ and $j$. A string
operator, denoted as $S^{s}_{ij}$, of overlapping $K^{\alpha}$
operators along this path $s$ can be represented as a site ordered
product of $\sigma^{\alpha}$ and $K^{\alpha,\beta}$ operators. We use the $K^{\alpha,\beta}$ notation in what follows when we wish to explicitly indicate the simultaneous operation of the constituent $\sigma^\alpha$ operators. If we assume that a $K^{\alpha,\beta}$ always acts first we see that the
total operator can be interpreted as creation of two vortex-pairs and
subsequent movement of one of the pairs along the path
$s$. Importantly, we see that $\sigma^{\alpha}$ correspond to a
rotation of one vortex-pair, whereas $K^{\alpha,\beta}$ moves the pair
without a rotation (see FIG. \ref{fig:movement}). If $i$ and $j$ are
neighboring sites and $s$ is a homologically trivial loop then by
definition $C_{(k,l)}=S_{ij}^s=\prod_p W_p$, see (\ref{eq:genloop}) and where the product is over all plaquettes enclosed by $s$. If we treat a vortex-pair as a
composite particle then the simplest loop operator $C_{(k,l)} = W_p$
(constructed from single $\sigma^\alpha$ operators) rotates the
composite particle by $2\pi$.  The resulting overall phase of
$e^{i\pi}$ suggests that the vortex-pairs are fermions for all values
of $J_\alpha$. 

Suppose now that the first and last links along the path $s$ are
$\alpha$- and $\mu$-links, respectively, and that the ends of the
string $S^s_{ij}$ are given by the operators $\sigma^\beta$ or
$K^{\beta,\alpha}$ and $\sigma^\nu$ or $K^{\nu,\mu}$.  Then
\be
\label{cfstring}
S_{ij}^{s} H S_{ij}^{s} = H + 2J_{\gamma} K_{ia}^{\gamma} + 2J_\tau
K_{jb}^{\tau}
\ee
where $a$ and $b$ are the sites connected to $i$ and
$j$ by the respective $\gamma$- and $\tau$-links, $\gamma \ne \alpha
\ne \beta$ and $\tau \ne \mu \ne \nu$. Taking the expectation value of both sides with respect to any translationally invariant state $\ket{\psi}$, which includes the vortex-free ground state \cite{Kit06,Lieb94}, we see that the expectation energy of the state $S^s_{ij} \ket{\psi}$ depends only on the ends of the string and this energy contribution is the same for links of the same type. This implies, even when
$J_x \ne J_y \ne J_z$ , that vortex-pairs created from the ground state can be propagated freely provided the relative orientation of each pair remains
constant. The expectation energy of the states created in this way can be calculated explicitly with respect to the ground state \cite{Kit06,Bas07}.

These fermionic vortex-pairs are distinct from the fermions
introduced as redundant degrees of freedom in \cite{Kit06}, those
obtained by Jordan-Wigner transformation \cite{Chen07a,
Chen07b,Feng07} and the vorticity preserving free-fermionic
excitations of \cite{Sch07}. In the gapped phase however,
the low-energy vortex-pair configurations are related to
certain fermionic $e$-$m$ composites of the toric code \cite{Kit06,Kit03}.
This last point is potentially relevant to the connection between the abelian and the non-abelian phases \cite{Woo08}.

%
%
%

The movement of vortex-pairs is in contrast to the situation
encountered when one wants to separate individual vortices. Crucially,
this cannot be done using overlapping $K^{\alpha}$ terms and indeed
can only be achieved if we use single $\sigma^\alpha$'s that do not,
in general, act on neighbouring sites. To this end we define
$D^{s}_{ij} \equiv \sigma_i^\alpha  \sigma_k^\beta....
\sigma_j^\gamma$, where it is understood that if $k$ and $l$ are
neighboring sites along some link, then $\alpha \ne \beta$. These operators satisfy
\begin{eqnarray} \label{vstring}
	D_{ij}^s H D_{ij}^s = E_0 + a J_x K^{x} + b J_y  K^{y}  + c J_z K^{z},
\end{eqnarray}
where $a+b+c=|s|$ for some integers $a$, $b$ and $c$ depending on the
path $s$. As before, suppose we take the expectation value of (\ref{vstring}) with respect to a translationally invariant state $\ket{\psi}$. In this case we see that the energy expectation value of the state $D_{ij}^s \ket{\psi}$ scales with $|s|$ and implies a string tension for states created in this way \footnote{This result does not imply vortex confinement because the state  $D_{ij}^s \ket{\psi}$ is  not necessarily the ground state of the relevant vortex configuration sector. Note that the perturbative results of \cite{Sch07} suggest the vortex configuration ground state energies, as a function of the distance between vortices, are bounded.}.

The above results, valid for all values of the parameters $J_\alpha$, are
in agreement with the perturbative analysis of the gapped abelian phase
\cite{Dus08}. There it was shown that while the repeated application of
single $\sigma^{\alpha}$ excites $e$ or $m$ toric code semions in the
low-energy dimerized subspace, it also introduces contributions to the
wavefunction from higher energy eigenstates. These high-energy
eigenstate contributions also occur when low energy vortex-pairs are
excited, but in this case two effective toric code
$e$-$m$ pairs are created in the effective system. An easy way to see this is to compare the expectation energy  $\langle K^{y,z} H K^{y,z} \rangle = E_0 + 4 J \langle K^{x}\rangle \approx E_0 + 2 J^2 /J_z $ , see \cite{Dus08}, to that of the ground state energy of the 4-vortex configuration $E \approx E_0 + J^4/2 J_z^3$, where $J=J_x=J_y$. However, since (\ref{cfstring}) implies that the vortex-pairs can be moved freely, there can be no increase in the contribution from these high energy
states as these pairs are propagated. This may be useful for the
experimental detection of anyons because, in the toric code, the $e$-$m$
pairs and single $e$ (or $m$) excitations have mutually anyonic
statistics. 

In conclusion, we have associated each of the loop symmetries of the full toroidal system with a term in the perturbative expansion. We then demonstrated the order at which the topological degeneracy is
broken and noted that, in the thermodynamic limit, it remains to all
orders. In a further analysis, we showed that the symmetries correspond to propagation
of vortex-pairs along closed loops. When treated as composite
particles, the vortex-pairs are fermions. We showed that these pairs
can be propagated with no additional energy cost but that, in general,
single vortices can not. In relation to the abelian phase we have included a discussion that shows how vortex-pairs maybe transported in way that keeps the fermionic population to a minimum. This argument relies on a detailed understanding of the supporting spectrum. It may be possible to make similar arguments in the non-abelian phase with inclusion of the effective magnetic field, however the necessary understanding of the supporting eigenspectrum is currently lacking. 

We acknowledge the support of SFI
through the PIYRA and of EU networks EMALI, SCALA, EPSRC, the
Royal Society, the Finnish Academy of Science and ICHEC. We thank Gregoire Misguich and Simon Trebst for discussions and
Timothy Stitt and Niall Moran for assistance in the early stages of this work.

\end{document}